\documentclass[manuscript]{aastex}
\usepackage{color}
\usepackage{graphicx}
\usepackage{multicol}
\usepackage[pagewise]{lineno}
\usepackage[normalem]{ulem}

\shorttitle{Dandelion Filament Eruption}
\shortauthors{Cabezas et al.}

\begin{document}

\title{``Dandelion'' Filament Eruption and Coronal Waves Associated 
with a Solar Flare on 2011 February 16}

\author{
Denis P. Cabezas\altaffilmark{1,2,3},
Lurdes M. Mart\'{i}nez\altaffilmark{2},
Yovanny J. Buleje\altaffilmark{2},
 Mutsumi Ishitsuka\altaffilmark{1},
Jos\'{e} K. Ishitsuka\altaffilmark{1},
Satoshi Morita\altaffilmark{4},
Ayumi Asai\altaffilmark{5},
Satoru UeNo\altaffilmark{3},
Takako T. Ishii\altaffilmark{3},
Reizaburo Kitai\altaffilmark{3,6},
Shinsuke Takasao\altaffilmark{3},
Yusuke Yoshinaga\altaffilmark{3,7}, 
Kenichi Otsuji\altaffilmark{3,4}, 
and
Kazunari Shibata\altaffilmark{3}}

\altaffiltext{1}{Geophysical Institute of Peru, 
Calle Badajoz 169, Mayorazgo IV Etapa, Ate Vitarte, Lima, Peru}
\altaffiltext{2}{Centro de Investigaci\'{o}n del Estudio de la Actividad
Solar y sus Efectos Sobre la Tierra, Facultad de Ciencias, Universidad
Nacional San Luis Gonzaga de Ica, Av. Los Maestros S/N, Ica, Peru}
\altaffiltext{3}{Kwasan and Hida Observatories, Kyoto University, 
Yamashina, Kyoto, 607-8471, Japan}
\altaffiltext{4}{National Astronomical Observatory of Japan, Osawa, Mitaka, 
Tokyo 181-8588, Japan}
\altaffiltext{5}{Unit of Synergetic Studies for Space, Kyoto University, 
Sakyo, Kyoto, 606-8502, Japan}
\altaffiltext{6}{Bukkyo University, 
Kita, Kyoto, 603-8301, Japan}
\altaffiltext{7}{Department of Astronomy, Kyoto University, 
Sakyo, Kyoto 606-8502, Japan}

\email{denis@kwasan.kyoto-u.ac.jp}

\begin{abstract}
Coronal disturbances associated with solar flares, such as H$\alpha$ Moreton waves, 
X-ray waves, and extreme ultraviolet (EUV) coronal waves are discussed herein in 
relation to magnetohydrodynamics fast-mode waves or shocks in the corona.
To understand the mechanism of coronal disturbances, full-disk solar observations 
with high spatial and temporal resolution over multiple wavelengths are of crucial 
importance.
We observed a filament eruption, whose shape is like a ``dandelion'',  associated 
with the M1.6 flare that occurred on 2011 February 16 in the H$\alpha$ images taken 
by the Flare Monitoring Telescope at Ica University, Peru. We derive the 
three-dimensional velocity field of the erupting filament.
We also identify winking filaments that are located far from the flare site in the 
H$\alpha$ images, whereas no Moreton wave is observed. By comparing the temporal 
evolution of the winking filaments with those of the coronal wave seen in the extreme 
ultraviolet images data taken by the Atmospheric Imaging Assembly on board the 
{\it Solar Dynamics Observatory} and by the Extreme Ultraviolet Imager on board the
{\it Solar Terrestrial Relations Observatory-Ahead}, we confirm that the winking 
filaments were activated by the EUV coronal wave. 
\end{abstract}


\section{Introduction}
Coronal disturbances associated with solar flares have been discussed in relation 
to magnetohydrodynamics (MHD) fast-mode waves or shocks in the corona \citep[and 
references therein]{pat2012,warm2015}. Moreton waves are flare-associated propagating 
wavelike features seen in H$\alpha$, especially in its wings \citep{moreton1960,
morram1960,athay1961}. This dynamic phenomenon is directional in restricted solid 
angles with arc-like fronts, and propagates away from the flare site at speeds of 
the order of 500 -- 1500~km~s$^{-1}$. 
\citet{uchi1968} proposed the following mechanism for producing Moreton waves: they 
are the intersection of MHD fast-mode shocks propagating in the corona with 
chromospheric plasma. A good correlation between H$\alpha$ Moreton wave and Type-II 
radio bursts \citep{kai1970} also supports this scenario.

Extreme ultraviolet (EUV) coronal waves are wavelike disturbances or expanding-dome 
structures observed in EUV passbands in the solar corona associated with flares. 
They were first observed by the Extreme-ultraviolet Imaging Telescope 
\citep[EIT;][]{delabou1995} on board the {\it Solar and Heliospheric Observatory}
\citep[{\it SOHO};][]{domingo95} and were called EIT waves \citep{thom1999,thom2000}.
Although EIT waves were expected to be the coronal counterpart of Moreton waves, the 
propagating speed was much lower (170 -- 350 km~s$^{-1}$) than that of Moreton waves 
\citep{klassen2000, Thom-Myers2009}. Therefore, the nature of EIT waves was questioned 
due the discrepancies on their physical characteristics as compared to Moreton waves. 
Simultaneous observations of EIT waves and H$\alpha$ Moreton waves \citep{eto2002} show 
that EIT wavefronts are not co-spatial with those of Moreton waves. On the other hand, 
\citet{warmuth2004a,warmuth2004b} and \citet{vrsnak2016} discuss that the velocity 
discrepancy can be explained by the deceleration of coronal waves.

``Fast'' EUV waves associated with flares have been observed by recent EUV observations 
made mainly by the Atmospheric Imaging Assembly \citep[AIA;][]{lemen2012} on board the 
{\it Solar Dynamics Observatory} \citep[{\it SDO};][]{pesnell2012} and by the
Extreme-Ultraviolet Imager (EUVI) from the Sun-Earth Connection Corona and Heliospheric 
Investigation \citep[SECCHI;][]{howard2008} on board the {\it Solar Terrestrial 
Relations Observatory} \citep[{\it STEREO};][]{kaiser2008}. Since the speed of the 
wavelike features is sometimes very fast and comparable to that of Moreton waves, these 
waves seem to be fast-mode MHD waves or shocks \citep[e.g.,][]{chen2011,asai2012}.
Recently, some authors linked the EUV waves with direct manifestations of CMEs rather 
than solar flares, but the exact physical mechanism by which they are related remains 
unclear \citep{biesecker2002,gallagher2011,nitta2013}.

Associated with flares and coronal waves, filaments and prominences at distances
are sometimes activated or excited into oscillation. This so-called ``winking'' 
(appearing and/or disappearing) motion of a filament is thought to be triggered by 
a Moreton wave \citep{smi1964}. \citet{eto2002} reported the simultaneous observation 
of an H$\alpha$ Moreton wave and a winking filament triggered by its association with 
a solar flare.
Winking filaments are more often observed for flares without Moreton waves, whereas 
Moreton waves are difficult to observe even for large flares. Winking filaments, 
therefore, constitute indirect evidence of coronal wave traveling in the corona 
\citep{shi2011}, which are called ``invisible'' Moreton waves \citep{smi1971}.
\citet{oka2004} on the other hand, reported a winking filament triggered by the passage 
of an EIT wave. \cite{taka2015} reported a winking activation in detail caused by an EUV 
coronal wave. \citet{warmuth2004b} discussed a schematic outline how to activate a winking 
filament by a Moreton wave and/or EIT wave. \citet{vrsnak2016}, furthermore, showed in 
their numerical simulation that the lateral expansion during an eruption is important for 
producing a Moreton wave.

In this paper we examine in detail the coronal disturbance features, such as filament 
eruption, winking filaments, and EUV waves associated with the M1.6 flare that occurred
on 2011 February 16. 
The EUV coronal wave of this flare has already been discussed \citep{harra2011,vero2011}.
In particular, these authors examined the spectroscopic features, based on the 
sit-and-stare spectral data in EUVs acquired by the EUV Imaging Spectrometer 
\citep[EIS;][]{culhane2007} on board {\it Hinode} \citep{kosugi2007}.
We use H$\alpha$ images taken by the Flare Monitoring Telescope \citep[FMT;][]{kurokawa1995} 
that was installed at National University San Luis Gonzaga of Ica, Peru.
We examine in particular the three-dimensional velocity field of the erupted filament
and the relation between the H$\alpha$ filament eruption and the EUV coronal wave.
We also examine the relation between winking features of H$\alpha$ filaments and the 
EUV coronal wave by analyzing observations from {\it SDO}/AIA and {\it STEREO-Ahead} 
({\it STEREO-A})/SECCHI/EUVI.
In Section 2 we describe the data used in this study. The analysis and results are 
presented in Section 3, and the conclusions are summarized in Section 4.

\section{Observations and Data}
\begin{figure}
\epsscale{0.80}
\plotone{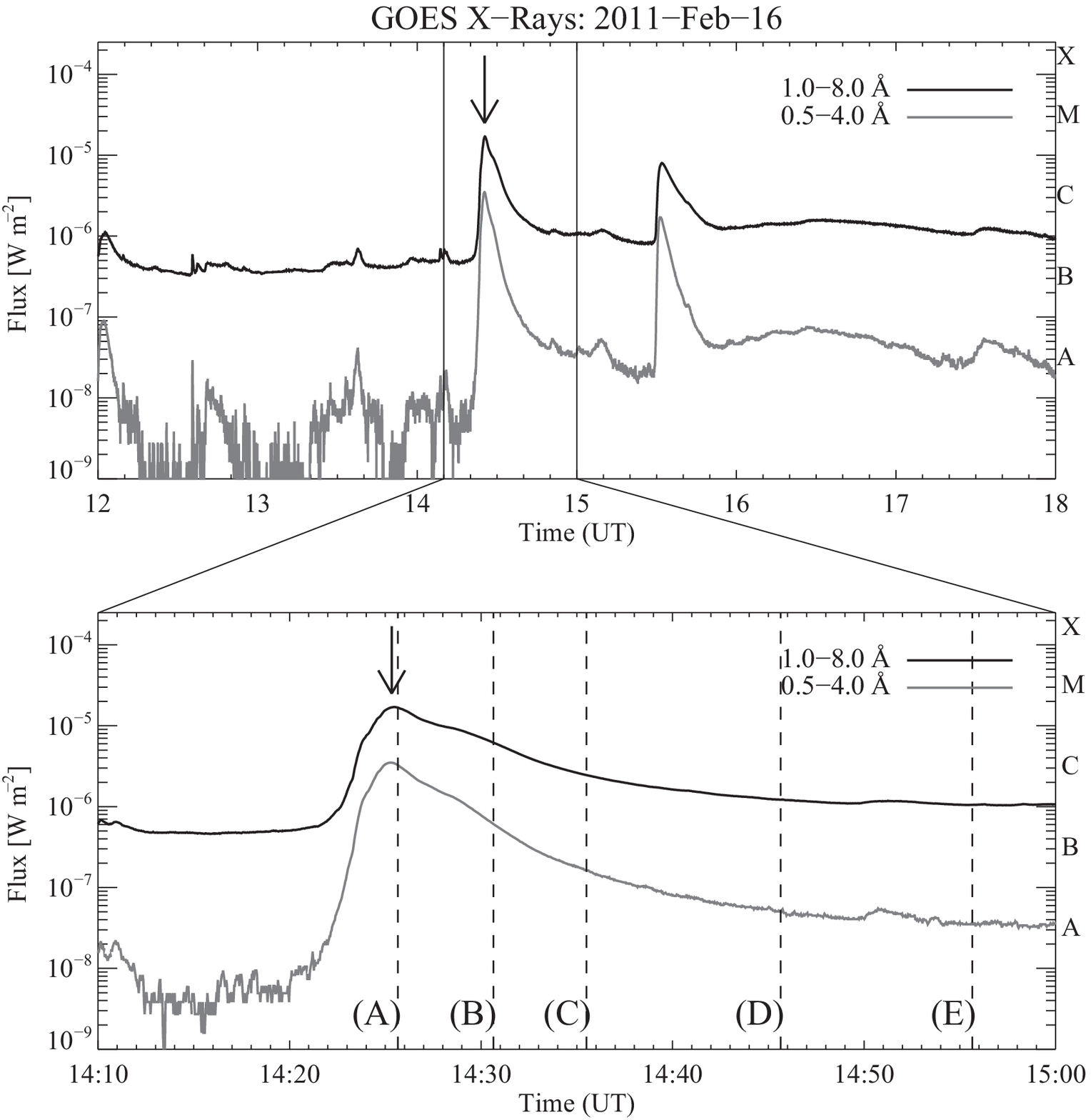}
\caption{
{\it GOES} X-ray emission at 1.0-8.0~{\AA} and 0.5-4.0~{\AA} channels of the 2011 February 
16 flare (with extended time range at the top and an enlarged view in the bottom panels). 
The arrows indicate the peak of the M1.6 flare at 14:25:20~UT, which occurred in the Active 
Region NOAA 11158. The second peak at 15:32~UT in the top panel corresponds to a C7.7 flare 
that occurred in the same active region. The vertical dashed lines in the bottom panel point 
out the times of the images shown in Figures~\ref{FIG5} and \ref{FIG6}.}
\label{FIG1}
\end{figure}

The flare in question occurred in the Active Region NOAA 11158 (S21$^{\circ}$, 
W30$^{\circ}$). It started at 14:19~UT and peaked at 14:25~UT on 2011 February 16, 
and was classified as M1.6 on the {\it GOES} scale (Figure~\ref{FIG1}). 
NOAA 11158 produced the first X-class flare in the solar cycle 24 during the 
previous day. Figure~\ref{FIG1} shows the time evolution of the {\it GOES} X-ray
emission for the February 16 flare.

For the analysis we use H$\alpha$ data taken by FMT that provide full-disk solar
images in the H$\alpha$ line center ($6562.8$~{\AA}), H$\alpha$ $-0.8$~{\AA}~(blue-wing),
H$\alpha$ $+0.8$~{\AA}~(red-wing), continuum, and limb prominence images (H$\alpha$ 
line center with occulting disk). 
The spatial resolution is about 2.0$^{\prime\prime}$, and the time cadence is 20 
seconds. Such multiwavelength full-disk observations with high time cadence are 
suitable for investigating the velocity field of erupting filaments \citep{morimoto2003a,
morimoto2003b,morimoto2010} and for detecting Moreton waves \citep{eto2002}. 
FMT was relocated in March 2010 from Hida Observatory, Kyoto University to National 
University San Luis Gonzaga of Ica in Peru by the international collaboration of the 
Continuous H-Alpha Imaging Network (CHAIN) project \citep{ueno2007} for coordinated 
solar observations. 

We cannot find any direct signatures of H$\alpha$ Moreton waves associated with the flare
not only in the FMT images (in H$\alpha$ line center and the $\pm0.8$~{\AA} wings) but 
also in their running-difference images.
However, we found indirect evidence of the coronal wave in the form of winking filaments 
that are located far from the flare site.
Figure~\ref{FIG2} shows a full-disk H$\alpha$ line center image of the flare taken by 
FMT. We show the flaring region with $f_{0}$ and the filaments are denoted $f_{1}$ to 
$f_{5}$. The boxes indicate the areas over which we calculate the intensities to 
identify their winking features activated by the flare and the coronal waves.
Associated with the flare, we also observe a filament eruption whose shape in H$\alpha$
images looks like a ``dandelion''. This feature is especially prominent in the red- 
and blue-wing images. Figure~\ref{FIG3} also presents FMT images of the flare, filament 
eruption, and winking filaments in H$\alpha$ $-0.8$~{\AA} (left), H$\alpha$ line center 
(middle), and H$\alpha$ $+0.8$~{\AA} (right) passbands.

The flare was observed in EUV passbands by {\it SDO}/AIA and {\it STEREO-A}/SECCHI/EUVI. 
AIA provides high-resolution full-disk images of the transition region and the corona in 
multiple wavelengths. We used AIA 171, 304, and mainly 193~{\AA} images to investigate 
the evolution of the flare, the associated erupting filament, and the coronal waves. 
The temporal resolution of the AIA 193~{\AA} data, which are attributed to a Fe~{\sc xii} 
line with the formation temperature of 1.25~MK, is 12 seconds, and the pixel size of the 
image is 0.60$^{\prime\prime}$.
At the time of the flare, {\it STEREO-A} was located 86.8$^{\circ}$ ahead of the Earth.
{\it STEREO-A}/EUVI images at 195~{\AA}, which again are mainly from the Fe~{\sc xii} line,
have a temporal resolution and pixel size of 5 minutes and 1.58$^{\prime\prime}$, 
respectively.

The flare was also captured in soft X-rays by the X-Ray Telescope \citep[XRT;][]{golub2007}
on board {\it Hinode}. XRT is sensitive to hot coronal plasma and can detect emissions
from plasma with temperatures of the order of 1 to several tens of MK. We used partial 
images with 2.5$^{\prime\prime}$ pixel resolution through the Thin-Be filter. 
In addition, to calculate potential magnetic field, we used the Potential Field Source 
Surface (PFSS) extrapolation model\footnote{http://www.lmsal.com/$^{\sim}$derosa/pfsspack/} 
\citep[][]{schrijver-derosa2003} that was applied to synoptic magnetograms taken 
by the Michelson Doppler Imager \citep[MDI;][]{scherrer95} on board {\it SOHO}, based on the 
LMSAL\footnote{Lockheed Martin Solar and Astrophysics Laboratory} forecaster program.

\section{Analysis and Results}
\subsection{Overview of active region and of flare}
Figure~\ref{FIG4} presents the temporal evolution of the flare in various wavelengths.
Columns from left to right are images in soft X-rays Thin-Be filter, in EUVs at 171 and
304~{\AA}, and in H$\alpha$ line center; and were taken by XRT, AIA, and FMT, respectively. 
Before the flare starts, the so-called ``sigmoid'' or {\it S}-shaped magnetic field 
structure \citep[e.g.,][]{rust1996,can1999} that lies in the east-west direction is 
observed in soft X-rays (see top left panel in Figure~\ref{FIG4}). 

During the time of the flare, NOAA~11158 was composed of several sunspots, as discussed 
in detail by \citet{tori2013} and \citet{sun2012}. In particular, two neighboring fast 
flux emergences ($P_{0}$ - $N_{0}$ and $P_{1}$ - $N_{1}$ pairs) appeared in line along 
the east-west direction and formed a quadrupolar distribution (see top panel of the 
H$\alpha$ $-0.8$~{\AA} column in Figure~\ref{FIG5}). Between the following negative spot 
$N_{0}$ of the western bipole and the preceding positive spot $P_{1}$ of the eastern bipole 
appeared a prominent filamentary structure along the magnetic neutral line on February 13, 
which developed for about three days.
The X-class flare on February 15 occurred along the elongated neutral line between
$N_{0}$ and $P_{1}$. This basic magnetic configuration remained even after the X-class 
flare and displayed the sigmoid structure seen in the XRT images.

The M-class flare in question, however, occurred at the eastern edge of the main neutral 
line between $N_{0}$ and $P_{1}$. When the impulsive phase of the flare just begins 
($\sim$14:22~{UT}), a small bright X-ray feature appears at the eastern-end of the sigmoid 
structure. The brightening has counterparts in H$\alpha$ and 304~{\AA} images, which are 
mainly from footpoints of the flaring loops. As of 14:31~{UT}, the filament eruption can be 
distinguished in EUV and H$\alpha$ images (see bottom panels of Figure~\ref{FIG4}).

The starting point of the filament eruption is also located at the eastern edge of the
neutral line between $N_{0}$ and $P_{1}$. This filament eruption is clearly discernible 
both in the H$\alpha$ and EUV images, as shown in Figure~\ref{FIG5}. It was activated near 
14:25~UT and displays an enlarging dark feature in H$\alpha$ blue-wing images, like a 
dandelion, whereas no manifestation is exhibited in the H$\alpha$ red wing until 14:32~{UT}.

\subsection{Eruption of Dandelion filament}
Here, we describe how we used H$\alpha$ images to derive the three-dimensional velocity 
field of the erupting filament.  
For the line-of-sight (LOS) velocity, we modified the method by \citet{morimoto2003a}. 
They computed the LOS velocities of erupting filaments observed with FMT by using the 
method based on the Beckers' cloud model \citep{beckers1964,mein1988}.
\citet{morimoto2003a} adopted the single cloud model. Because FMT has only three 
(center and $\pm0.8$~{\AA}) wavelength data points for the four unknowns (source function, 
optical depth, Doppler velocity, and Doppler width) of a cloud, they further assumed a 
fixed Doppler width in their calculation. 
Our method, on the other hand, deals even with the Doppler width as a free parameter.
This can be implemented as follows: in the quiescent phase of the filament, we derived
the parameters assuming the value of the Doppler width, as was done in \citet{morimoto2003a}.
To analyze a time series, we search for the best-fit solution around the quiescent
parameter values found in the initial time step, and select the ``nearest'' local minimum
in the parameter field. This selection of the nearest values is based on the assumption
that the physical parameters in a single cloud do not change so quickly.
Next, we search the parameters in the next time step around the parameters derived in 
the previous time step, and so on.
We did not set a ``pre-defined range'' in our method, we used the hybrid method for 
the fitting \citep{powell1970}.
The error of the derived LOS velocity is roughly comparable with the errors obtained 
with the method of \citet{morimoto2003a} and is about $\pm$10~km~s$^{-1}$, 
while the derived LOS velocity tends to be overestimated for the fixed Doppler width,
since the Doppler width is thought to be larger for an erupting filament. The error
could be, however, a little bit worse due to misalignment of the wavelengths of the FMT 
filters from the nominal values. Comparing the wing data, the wavelength of the red wing 
($+0.8$~{\AA}) seems to be set closer to the line center than that of the blue wing 
($-0.8$~{\AA}). The error of the LOS velocity, therefore, is roughly $\pm$15~km~s$^{-1}$.

The left column in Figure~\ref{FIG6} shows the derived temporal evolution maps of the LOS 
velocity. The colors show blueshift (i.e., moving toward us) and redshift (i.e., moving 
away from us) of the filament. Near 14:30~UT, the filament begins to erupt northward and 
simultaneously shows blueshift features, especially at the leading part, with a LOS velocity 
of about $-30$~km~s$^{-1}$. At the root of the filament, in contrast, it shows a redshift 
with a velocity of about 20~km~s$^{-1}$.
The filament eruption is clearly seen until 14:35~UT. At this time, the LOS velocity 
at the leading edge is nearly $-25$~km~s$^{-1}$. Next, the front edge gradually becomes 
invisible at 14:45~UT, whereas near 14:55~UT, the filament shows a dominant redshift 
pattern.
The observed blueshift velocity of about $-30$~km~s$^{-1}$ is slightly larger than the 
result by \citet{vero2011}, while the temporal evolution is roughly consistent.

We confirm the temporal LOS motion with the side views of the filament eruption observed 
from {\it STEREO-A}, since it was located at 86.8$^{\circ}$ ahead of the Earth. In the 
EUV 195~{\AA} images of {\it STEREO-A}/SECCHI/EUVI, we see  movement of the cold plasma 
as dark features (second column in Figure~\ref{FIG5}). 
The dark feature is ejected from the flare site, and the traveling velocity of the front 
(i.e., the fastest) part is about 280~km~s$^{-1}$. The horizontal component of the velocity 
(i.e., the velocity in the direction from the Sun to the Earth) is about 230~km~s$^{-1}$. 
FMT cannot detect clouds with velocity greater than 50~km~s$^{-1}$, because the wavelengths 
of the wing data are set at $\pm$0.8~{\AA}. Therefore, FMT missed the front part of the 
ejection.

By comparing the EUVI 195~{\AA} and FMT H$\alpha$ $-0.8$~{\AA} images (Figure~\ref{FIG5}), 
the H$\alpha$ filament seems to correspond to the main (i.e., the darkest) part of the 
filament in the 195~{\AA} images. The main part of the filament has a velocity of about 
110~km~s$^{-1}$, and the horizontal component has a velocity of about 90~km~s$^{-1}$. 
The velocity is much larger than the LOS velocity derived by FMT (30~km~s$^{-1}$). LOS 
velocities derived from the cloud model reflect representative values of moving clouds, 
and the fine LOS velocity field in the clouds could be lost. 
Besides, the velocity in the Earth-Sun direction is derived by following dark features seen 
in EUVI images, and is not derived by tracing the moving ejecta. Alternatively, we could 
detect a slower component of the erupting filament, because the velocity of about 
90~km~s$^{-1}$ far exceeds the detection limit of FMT.

For the tangential velocity of the erupting filament (that is, the velocity of the 
filament in the plane of the sky), we traced the apparent motion of filament blobs by 
using the local correlation tracking (LCT) method. The middle column in Figure~\ref{FIG6} 
shows the temporal evolution of the erupting filament with arrows that show the orientation 
and magnitude of the tangential velocity.
The filament blobs are moving upward (i.e., northward) with a tangential velocity of 
about $10$~km~s$^{-1}$. The dark feature seen in EUVI 195~{\AA} images has a northward 
velocity of about 65~km~s$^{-1}$, which is faster than the tangential velocity derived 
from FMT. This result is explained by the excessively large mesh size used for the LCT 
method, which means that features moving at such high speed could not be detected even 
if they were present. The seeing condition prevents us from reducing the mesh size for 
the FMT images. Therefore, the derived tangential velocity is a somewhat averaged value 
of the main part of the ejected filament. 

In the right column of Figure~\ref{FIG6}, we show the temporal evolution of the inclination
of the ejecta. The inclination angle is set to zero for the upward direction normal to 
the solar surface.
In the early phase of the ejection (around 14:30~UT), the direction is nearly horizontal
with respect to the solar surface, with a slight upward inclination. During the main phase 
of the ejection, which occurs around 14:35~UT, the direction is still nearly horizontal but 
with a slight downward inclination. Next, the inclination distribution becomes complex showing 
upward and downward motions, probably because of the coexistence of upward and downward moving 
features in the later phase. 
Although the error in the velocity measurements is probably large, we believe that the
ejecta move nearly horizontally with respect to the solar surface.
Further, in Figure~\ref{FIG5} (EUVI panel at 14:30~UT), we depict the plane 
of the solar surface (yellow arrows over the heliographic coordinates) and a vertical 
component perpendicular to the surface (arrow labeled ``vertical"). The dashed 
green arrow highlights the direction of the filament eruption as seen by {\it STEREO-A}, 
and $\varphi$ is the inclination angle with respect to the vertical on the solar surface.
From EUV observations (193~{\AA} and 195~{\AA}) we infer the geometrical components
of the erupting material, which led us to estimate $\varphi$~$\approx$~58$^{\circ}$.
So that, the filament in EUVI was ejected having an elevation angle of about 32$^{\circ}$.

\subsection{Filament oscillations and EUV waves}
\begin{figure}
\epsscale{0.75}
\plotone{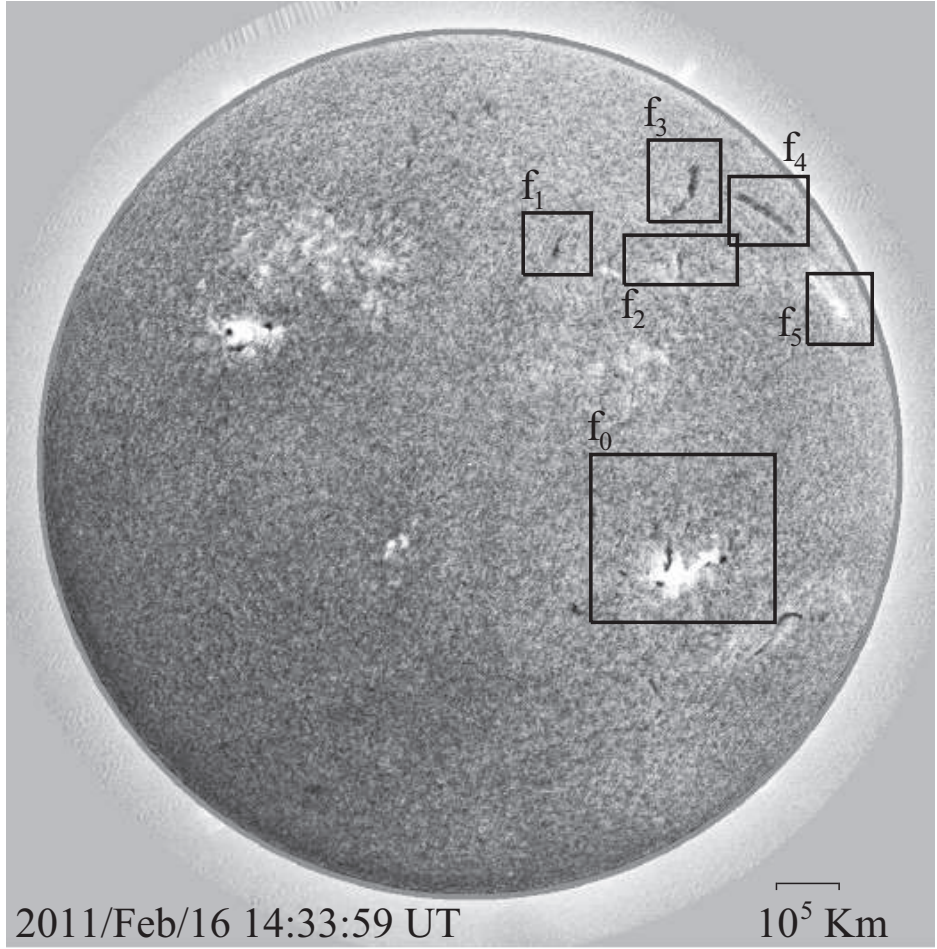}
\caption{
Full disk solar image (solar north is up and west is to the right) of the flare in H$\alpha$ 
line center taken by the FMT at National University San Luis Gonzaga of Ica, Peru. 
The box $f_{0}$ shows the flaring region, whereas the fields of view $f_{1}$ to $f_{5}$ contain 
the quiescent filaments. Filaments $f_{2}$ and $f_{5}$ are analyzed in more detail in 
Figure~\ref{FIG7}.}
\label{FIG2}
\end{figure}

\begin{figure}
\epsscale{0.85}
\plotone{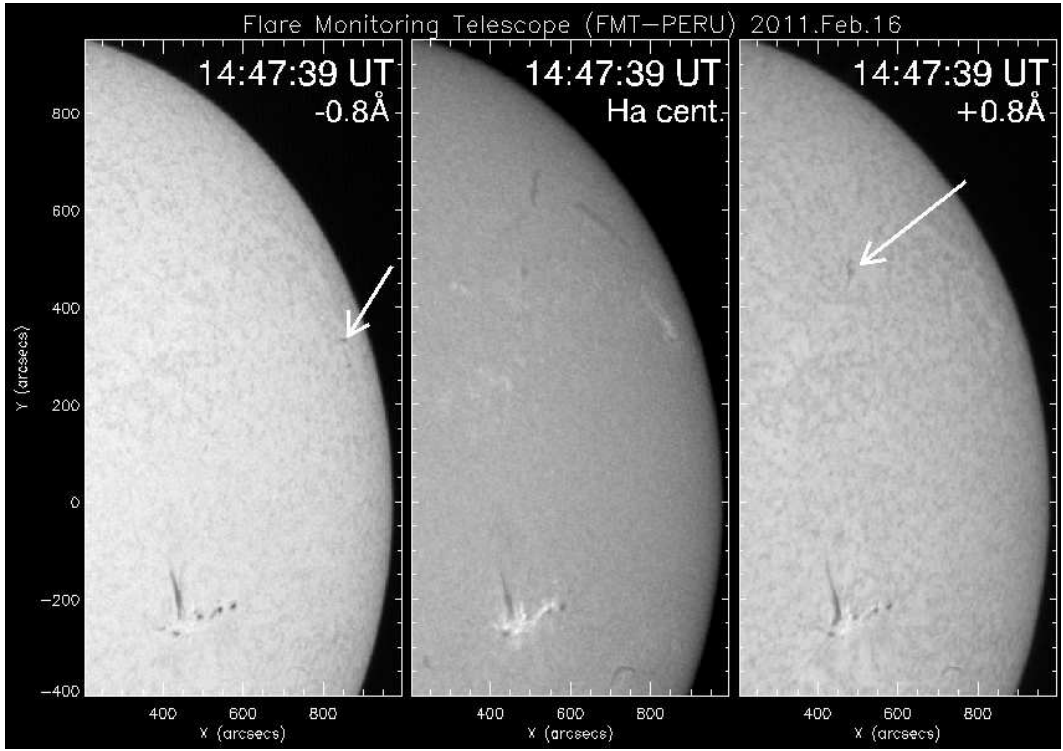}
\caption{Temporal evolution of the filament eruption and the oscillating filaments 
(arrows), associated with the 2011 February 16 flare. H$\alpha$ $-0.8$\AA~(left panel), 
H$\alpha$ line center (middle panel), and H$\alpha$ $+0.8$\AA~(right panel) images taken by 
FMT are shown. [See the electric edition of the Journal for an associated animation].}
\label{FIG3}
\end{figure}

\begin{figure}
\epsscale{0.80}
\plotone{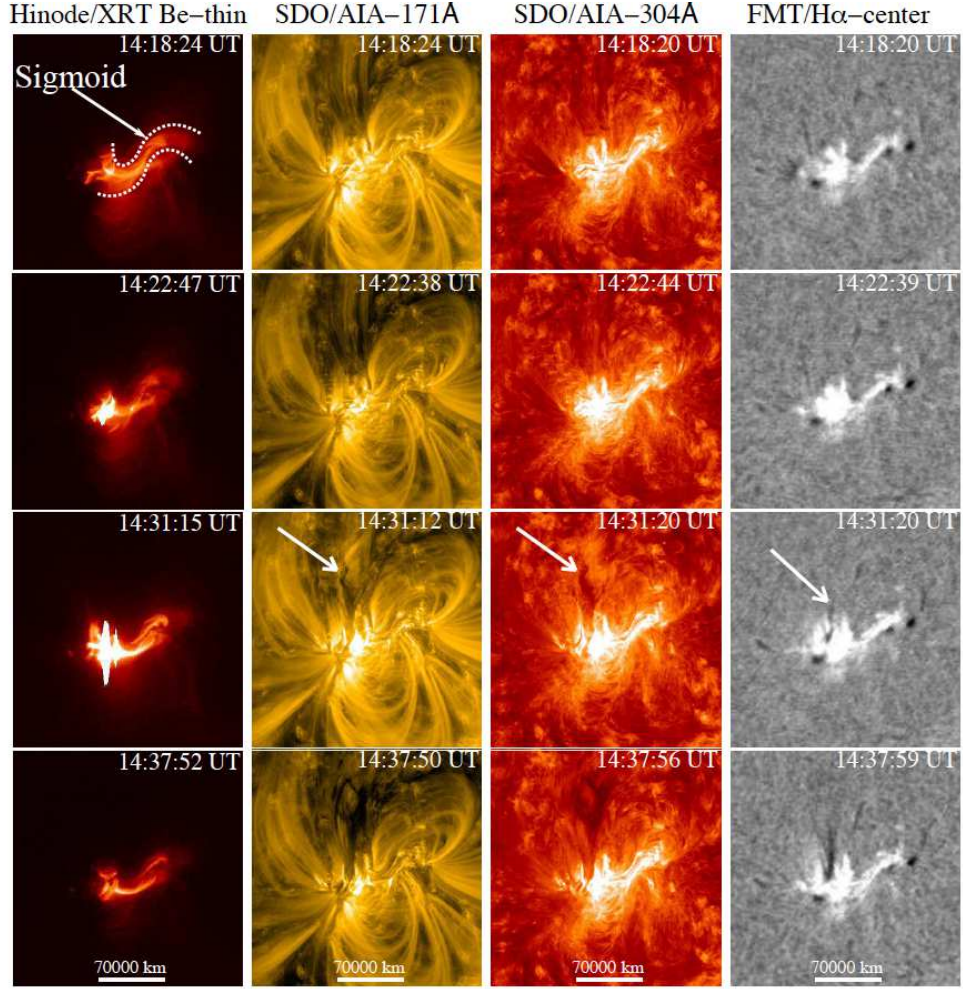}
\caption{
Multiwavelength time sequences of the 2011 February 16 flare. Columns from left to right 
are {\it Hinode} XRT Thin-Be broadband filter, {\it SDO}/AIA at 171 and 304~{\AA}, and FMT 
H$\alpha$ line center images. At 14:18~{UT} a sigmoid structure is clearly observed in 
X-ray images, the dotted lines outline the sigmoidal shape prior to the flare.
In 171~{\AA}, 304~{\AA}, and H$\alpha$ images a dark feature (filament eruption) becomes 
visible as of 14:31~{UT}. We pointed out the dark features by arrows.}
\label{FIG4}
\end{figure}

\begin{figure}
\epsscale{0.83}
\plotone{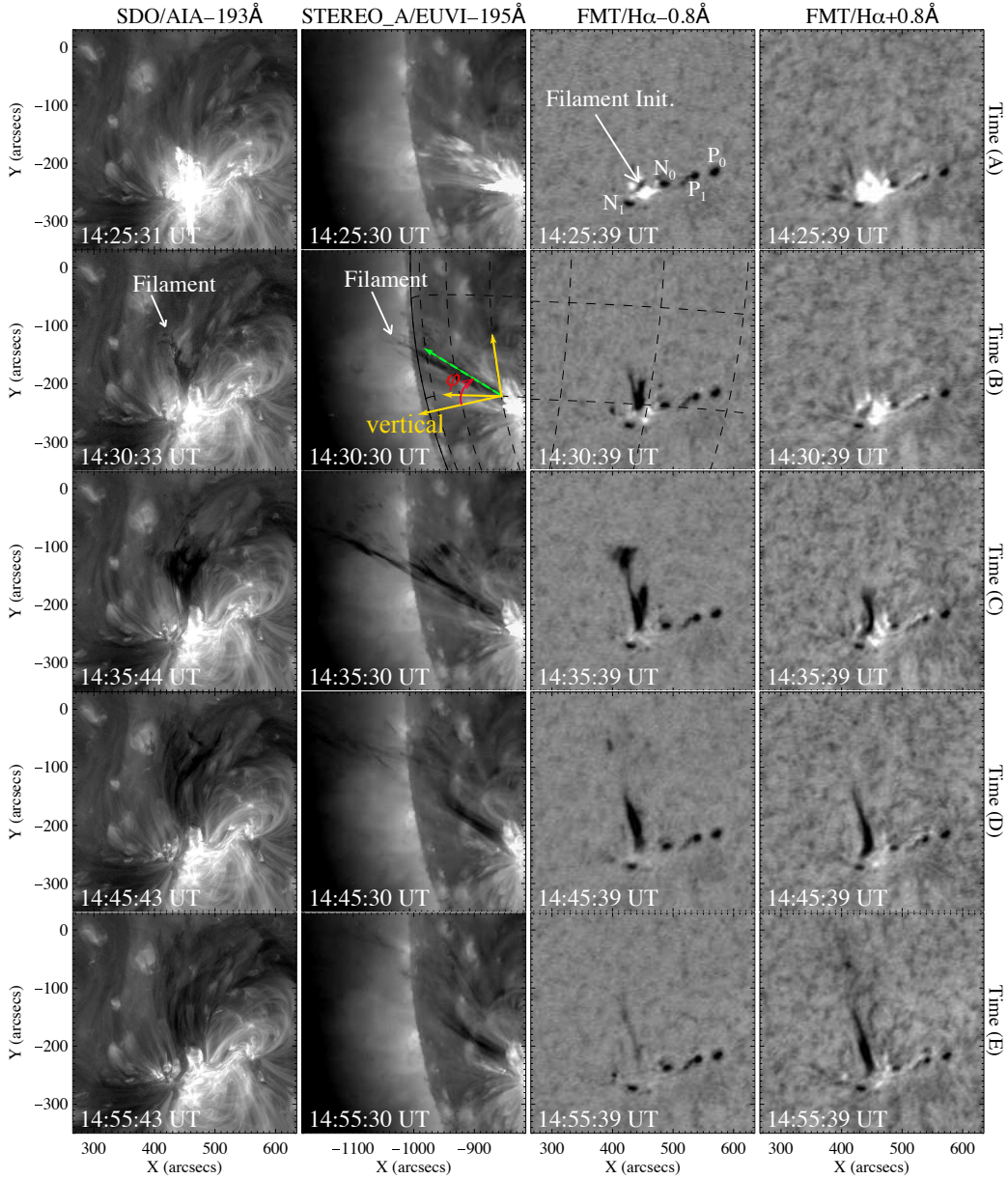}
\caption{
Temporal evolution of filament eruption. From left to right: AIA 193~{\AA}, EUVI 
195~{\AA}, and FMT H$\alpha$ $-0.8$ and $+0.8$~{\AA} images. The {\it STEREO-A}/SECCHI/EUVI 
images show the side view of the filament, since it was located 86.8$^{\circ}$ ahead of the 
Earth. The dashed green arrow at 14:30~UT point out the direction of the ejecta 
and the yellow arrows constitute the projected plane on the solar surface, $\varphi$ is the 
inclination angle with respect to the vertical component normal from the solar surface.
In the top panel of the third column, the arrow indicates the location of the 
erupting filament, whereas $P_{0}$, $N_{0}$, $P_{1}$, and $N_{1}$ denote the 
sunspot distributions.}
\label{FIG5}
\end{figure}

\begin{figure}
\epsscale{0.70}
\plotone{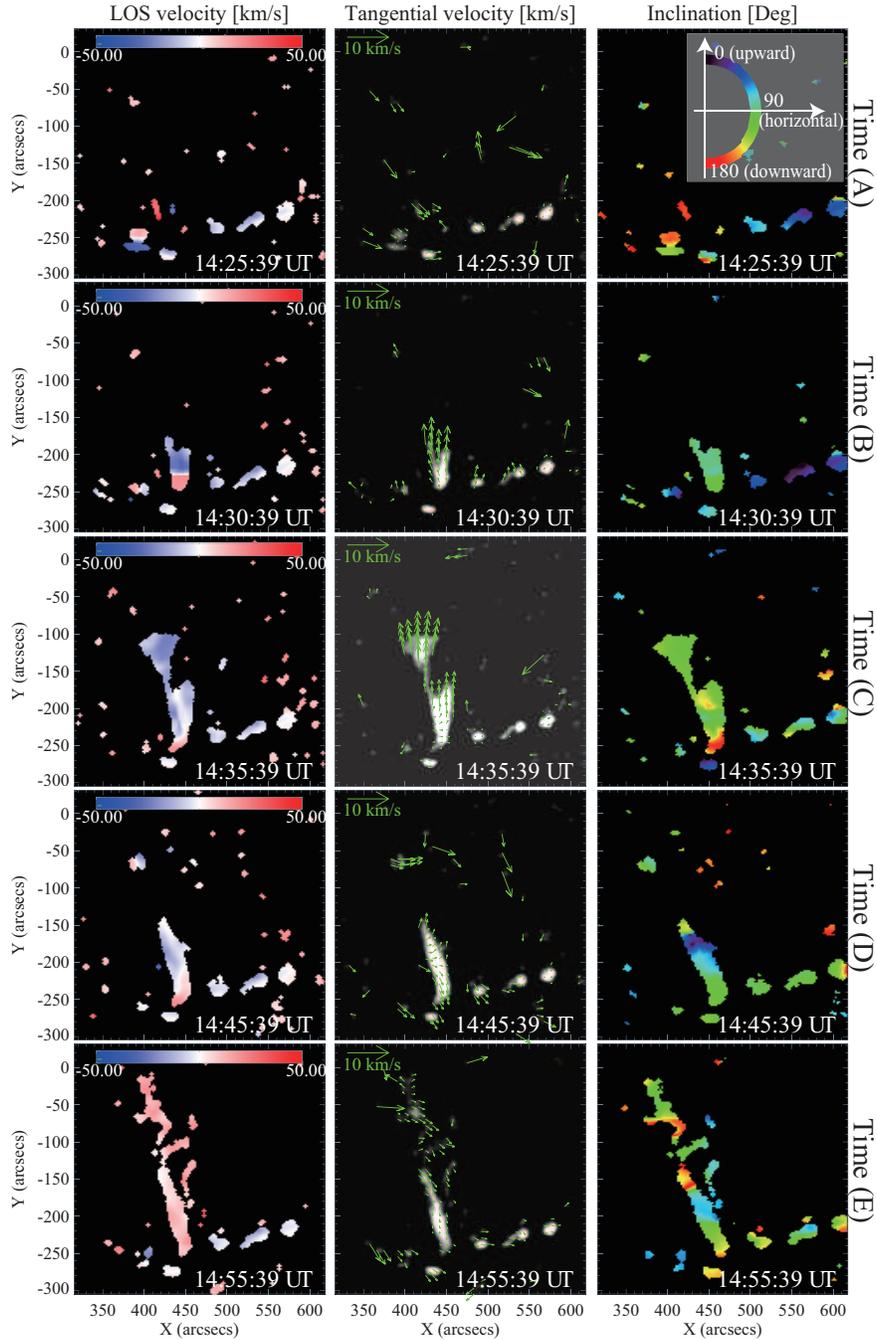}
\caption{
Line-of-sight (LOS) velocity (left column), tangential velocity (middle column), 
and inclination (left column) maps of the erupting filament. 
The LOS velocity was obtained by calculating the H$\alpha$ profile based on the cloud model.
The tangential velocity was calculated by tracing the motion of the internal structures 
in successive images.
The inclination is the result of the coordinate transformation of the velocity vectors and 
corresponds the top view of the filament, which is normal to the solar surface.}
\label{FIG6}
\end{figure}

In the H$\alpha$ data, we also observe winking (activation and/or oscillation) of the 
filaments associated with the flare. As we see in Figure~\ref{FIG2}, some filaments are 
labeled $f_{1}$ to $f_{5}$. These filaments are located far from the flare site. We 
identify a clear winking feature for filament $f_{2}$ and a much fainter signature for 
filament $f_{5}$ (see also the Figure~\ref{FIG3}).

\begin{figure}
\epsscale{0.85}
\plotone{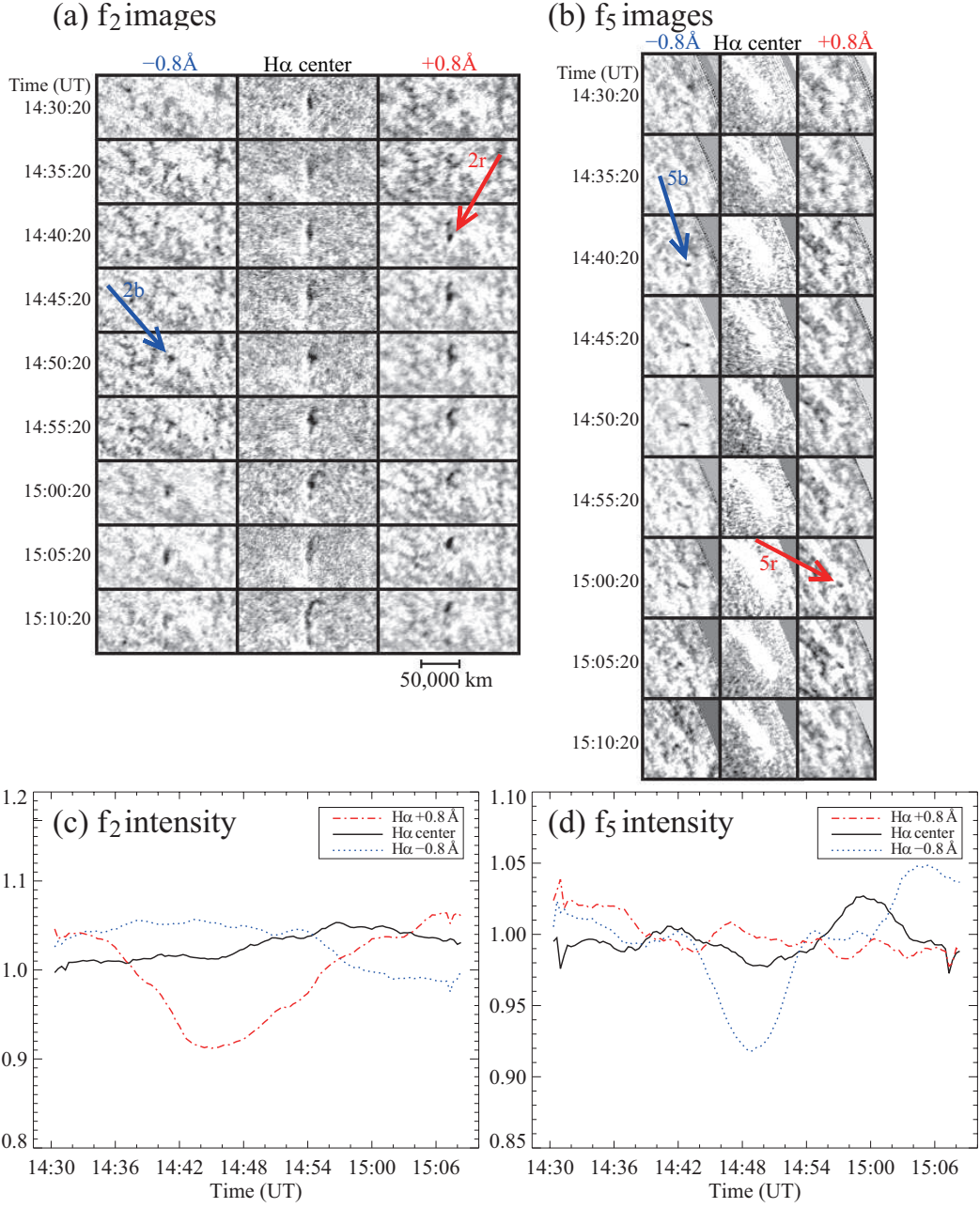}
\caption{
(a), (b) H$\alpha$ images: H$\alpha$ $-0.8$~{\AA} (left), H$\alpha$ center (middle),
and H$\alpha$ $+0.8$~{\AA} (right) of the oscillating filaments $f_{2}$ and $f_{5}$, 
respectively. The field of view is shown in Figure \ref{FIG2}. 
The arrows indicate the starting time of the oscillation both in H$\alpha$ $+0.8$~{\AA} 
(labeled $2r$ and $5r$) and H$\alpha$ $-0.8$~{\AA} ($2b$ and $5b$).
(c), (d) Time profiles of intensities for regions $f_{2}$ and $f_{5}$, respectively.
The horizontal axis is the time in UT, and the vertical axis is the normalized intensity.
The solid, dotted, and dash-dot lines are the intensities in H$\alpha$ center, $-0.8$~{\AA}, 
and $+0.8$~{\AA}, respectively.}
\label{FIG7}
\end{figure}

The time series of the winking filament $f_{2}$ is displayed in Figure~\ref{FIG7}(a)
at H$\alpha$ $-0.8$~{\AA} (left), H$\alpha$ center (middle), and H$\alpha$ $+0.8$~{\AA} 
(right) wavelengths. The field of view of the $f_{2}$ region is the same as in 
Figure~\ref{FIG2}. At 14:40~UT, the filament starts winking. First, a dark feature becomes 
prominent in the red-wing images, as shown by the arrow labeled $2r$ in Figure~\ref{FIG7}(a). 
The feature is dominant until 15:05~UT, and then it fades slowly away. However, at about 
14:50~UT, a dark structure starts to appear in the blue-wing images, as pointed out by 
another arrow ($2b$). This structure is noticeable until 15:10~{UT}, when it disappears. 
In the H$\alpha$ center, in contrast, no significant changes are observed.

Figure~\ref{FIG7}(c) plots the temporal variations of the intensities calculated for the 
$f_{2}$ region in the three wavelengths. In the red-wing plot (dash-dotted line), we 
clearly identify a decrease of the intensity associated with the appearance of the dark 
feature in the red-wing images.
The decrease starts at about 14:36~UT and reaches maximal depletion around 14:45~UT, after 
which it recovers. The signal in the blue-wing plot (dotted line), however, remains roughly 
constant until 14:55~UT and then starts to decrease gradually. In the H$\alpha$ center 
plot (solid line), the signal slightly increases.

A similar analysis for filament $f_{5}$ is presented in Figures~\ref{FIG7}(b) and (d). 
Unlike $f_{2}$, the dark structure appears mainly in the blue-wing~(left panels).
The dark feature is evident from 14:40~UT until 15:05~UT. Conversely, at 15:00~UT, 
we recognize a very weak feature in the red-wing (see arrow labeled $5r$ in 
Figure~\ref{FIG7}(b)).
The temporal variation of the intensities is plotted in Figure~\ref{FIG7}(d). 
The winking feature dominates during the darkening of the blue-wing plot (dotted line), 
whereas the red-wing (dash-dotted line) and H$\alpha$ center (solid line) plots show 
gradual variation and no clear signals caused by filament oscillation. For the other 
filaments $f_{1}$, $f_{3}$, and $f_{4}$, we verified the intensity profiles and found 
no winking patterns.

We also examine the temporal evolution of the coronal waves associated with the 2011 
February 16 flare and the relation with the oscillating filaments.
The time sequences of the EUV waves produced by the flare are presented in the top 
panels of Figure~\ref{FIG8} (panels a~--~d). These are the AIA~193~{\AA} running 
difference (20-minutes difference) images. The white arrows highlight the moving wavefronts, 
whereas the locations of the oscillating filaments $f_{2}$ and $f_{5}$ are marked by 
the boxes.
The distances of the filaments ($f_{2}$ and $f_{5}$) measured from the flare site along 
the spherical solar surface are $\approx$5.1 $\times$ 10$^{5}$ and $\approx$6.4 $\times$ 
10$^{5}$~km, respectively. We derive the speeds of the EUV wave by following the leading 
edge of wavefront consecutively along the spherical lines to the filaments $f_{2}$ 
(path 2) and $f_{5}$ (path 1) directions. The mean velocities computed from a linear 
fit are 430~${\pm}$~29~km~s$^{-1}$ and 672~${\pm}$~24~km~s$^{-1}$, respectively.
The EUV wave have been studied by several authors \citep[][]{harra2011,vero2011,
long2013,nitta2013}, and the propagating speed ranges from 500 to 700~km~s$^{-1}$,
which changes depending on the propagating direction. Therefore, the velocities
derived in this study are comparable with those results.

At 14:28~UT (Figure~\ref{FIG8}a), a well-defined wavefront propagates mainly toward the 
north, and a bright erupting material is observed that corresponds to the dandelion filament. 
At about 14:31~UT (Figure~\ref{FIG8}b), the central part of the wavefront seems to stop 
and to exhibit a stronger amplitude: this occurs just when the wave is approaching the 
edges of a weak active region that is located north of NOAA 11158. The observed enhanced 
pattern and the changes in the wavefront progression are tentatively attributed to the 
compression of plasma generated by the interaction between the coronal wave and the confined 
magnetic system of the active region \citep{uchi1974}. 
We plot the potential magnetic field configuration derived by {\it SOHO}/MDI in 
Figure~\ref{FIG8}(e). The red triangle indicates the location of the northern weak active 
region. We confirm that the propagating wavefront tends to stop there. Note that, at 
14:35~UT (Figure~\ref{FIG8}c), the leading part of the wavefront decelerates, perhaps 
retained by the magnetic field configuration, whereas the west part is slightly refracted 
and continues traveling northwest (Figure~\ref{FIG8}d). 

Figure~\ref{FIG8}(f) shows the progression of the most prominent fronts of the EUV wave from 
14:27 to 14:40~UT, the positions of the weak active region, and the oscillating filaments 
$f_{2}$ and $f_{5}$. At 14:37~UT, the wavefront labeled 5 approaches $f_{5}$ and, at 14:40~UT, 
wavefront 6 reaches $f_{2}$. These times are consistent with those identified both in the 
images and in the intensity plots of $f_{2}$ and $f_{5}$ (Figure~\ref{FIG7}).
Figure~\ref{FIG9} presents the time-distance diagram of these temporal behaviors.
We measured the temporal evolution of the wavefronts along the paths, as shown by 
the projected dashed lines in Figure~\ref{FIG8}(f).
The squares $\Box$ and circles $\circ$ plotted together with linear fit (dotted lines) in 
Figure~\ref{FIG9} show the times and the positions of the wavefronts as they move along 
paths 1 and 2, respectively. Diamonds $\Diamond$ indicate the oscillating filaments 
$f_{5}$ and $f_{2}$, which are on straight lines, as shown by the dotted lines. Therefore, 
the observed winking filaments are assumed to have been triggered by the passage of the 
EUV coronal wave. 

\section{Summary and Conclusions}
We observed a filament eruption shaped like a dandelion and associated with the 
flare that occurred at NOAA~{11158}. We analyze the H$\alpha$ images taken with 
FMT in Peru and derive the three-dimensional velocity field by using a new method.
To derive the LOS velocity, we use the modified cloud model that was originally 
developed for the FMT data by \citet{morimoto2003a}.
We compare the temporal behavior of the erupting dandelion filament with its EUV
side views observed by {\it STEREO-A}/SECCHI/EUVI. We realize that the LOS velocity
derived from H$\alpha$ images of FMT gives representative values of the moving 
clouds and that the very fast components seen in the {\it STEREO-A}/EUVI images 
are probably lost due to the detection limit of FMT.
The tangential velocity of the dandelion filament is derived by applying the local 
correlation tracking (LCT) method to the H$\alpha$ images. We derive a representative 
(averaged) tangential velocity of the main part of the filament, although we could 
not capture small fast-moving features, if they exist, because of the limited mesh 
size of the LCT method.
By combining these LOS and tangential velocities, and applying a coordinate 
transformation, we also derive the inclination of its velocity vectors. We confirm 
that the dandelion filament is ejected nearly horizontally with respect to the solar 
surface.

\begin{figure}
\epsscale{1.0}
\plotone{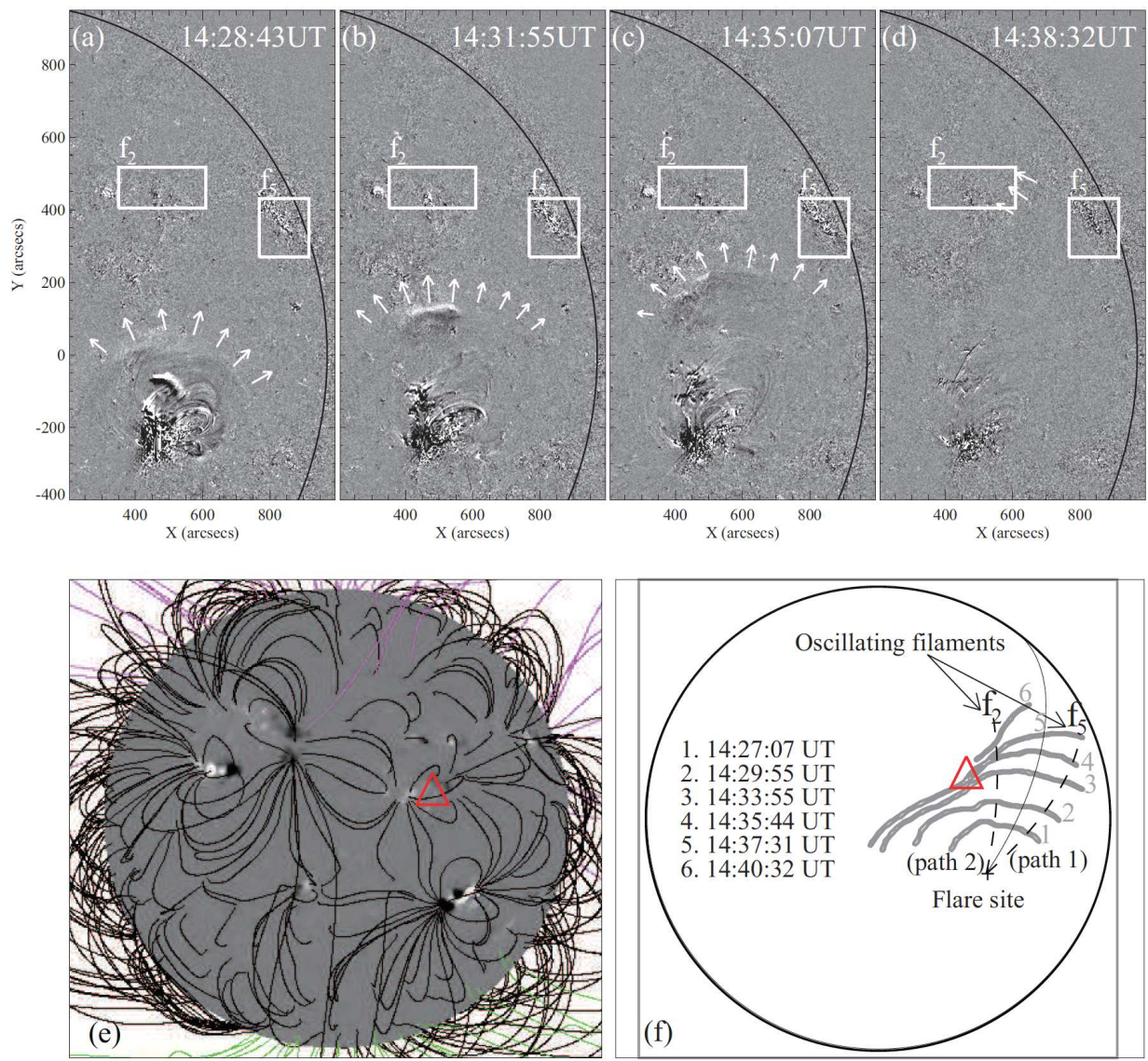}
\caption{
(a)~--~(d) Time sequences of running difference images of the {\it SDO}/AIA 193~{\AA} images.
The regions of the H$\alpha$ filaments $f_{2}$ and $f_{5}$ are marked by the rectangles.
(e) Potential magnetic field configuration extrapolated from the {\it SOHO}/MDI magnetogram 
based on the PFSS model. The red triangle denotes the location of the weak active region. 
(f): A schematic illustration of the coronal wave progression and its interaction with 
the filaments $f_{2}$ and $f_{5}$. The gray strokes outline the wavefronts at the given time 
and the arrows indicate the filaments position.}
\label{FIG8}
\end{figure}

\begin{figure}
\epsscale{0.85}
\plotone{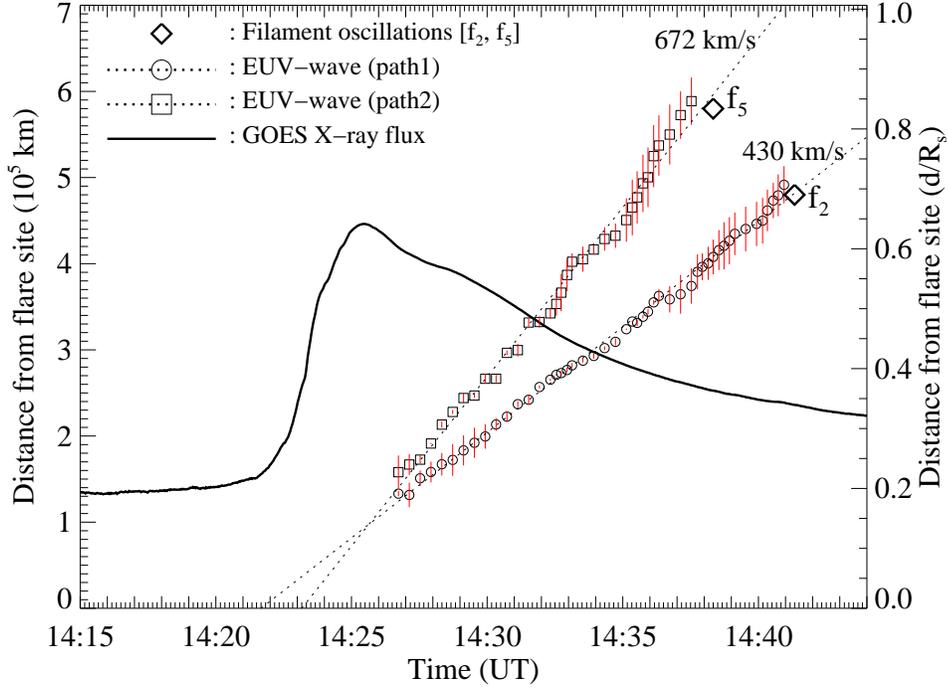}
\caption{
Time-distance diagram of EUV wave and oscillating filaments $f_{2}$ and $f_{5}$, 
measured along the paths shown by dashed lines in Figure~\ref{FIG8}(f).
The diamonds $\Diamond$ mark the positions of filaments $f_{2}$ and $f_{5}$, whereas the 
wavefront distances at given time are represented by $\circ$ and $\Box$ together with the 
standard deviation (error bars) along paths 1 and 2, respectively. R$_{s}$ is the solar 
radius ($\approx$695 800~km). For comparison, we also plot the {\it GOES} X-ray flux in 
the 1.0-8.0~{\AA} channel (solid line).}
\label{FIG9}
\end{figure}

We also observe the winking of H$\alpha$ filaments that are located far from the 
flare site, and examine the relation between the winking filaments and the EUV 
coronal wave associated with the flare.
As summarized in Figures~\ref{FIG8} and \ref{FIG9}, H$\alpha$ filaments $f_{2}$ 
and $f_{5}$ start oscillating when the EUV coronal wave arrives, so we confirm that
the interaction between the EUV wave and the H$\alpha$ filaments causes the filaments
to wink.

The derived traveling speed of the observed EUV waves is about 430 to 672~km~s$^{-1}$. 
Associated with this flare, Type II radio burst was also observed \citep{harra2011,
nitta2013}. These results are a strong indication that the observed EUV waves are MHD 
fast-mode wave or shock. 
Moreover, we observed the filament eruption ejected tangentially to the solar 
surface. This seems to be suitable to generate a H$\alpha$ Moreton wave, because the 
wavefront could easily make contact with the chromospheric plasma, even if a shock front 
is generated only in some restricted range at the front of the ejected filament.
Nevertheless, we do not observe the Moreton wave associated with the flare. This is probably 
because the downward velocity of the wavefront is small (about 20~km~s$^{-1}$) toward the 
solar surface, as reported by \citet{harra2011} and \citet{vero2011}, and probably because 
the EUV coronal wave (or shock) is not strong enough to generate an observable Moreton wave 
in the H$\alpha$ band.

We derived the three-dimensional velocity field of the filament eruption. We could follow 
the temporal evolutions of the velocity field and inclination of the ejecta. We also examined 
the temporal behaviors of the EUV waves associated with the filament eruption. These provide 
us an overall picture on the development of the EUV coronal wave and a filament ejection. 
Although FMT may have missed to capture very fast components the close correlations in time 
and main direction of the EUV wave progression with those of the filament eruption, leads us 
to conclude that the filament is the driver of the observed EUV wave.
Although the work we did by measuring the three-dimensional velocity fields of erupting 
filament and examining the relation between velocity fields and coronal EUV wave 
have shown a direct link among both phenomena, further studies are needed to fully understand 
filament eruptions as a driver of coronal EUV waves and coronal mass ejections (CMEs). 
Statistical studies to derive the tendency of filament eruptions to generate coronal waves 
and/or CMEs could be useful for this.
%

\acknowledgments
We are grateful to the anonymous referee for helping us to clarify and improve this 
manuscript.
DPC expresses special thanks to Dr. Mutsumi Ishitsuka for motivation and an invitation 
into the world of solar physics research, and is very grateful to all the staff members 
of the Kwasan and Hida observatories of Japan for all of the support and discussions 
during the FMT workshops and working-group meetings conducted in Japan and in Peru.
The authors are also grateful to the SDO/AIA and STEREO/EUVI teams for providing the high 
quality data used in this study. {\it Hinode} is a Japanese mission developed, launched, 
and operated by ISAS/JAXA, in partnership with NAOJ, NASA and STFC (UK). Additional 
operational support is provided by ESA and NSC (Norway).
This work was supported by JSPS KAKENHI Grant Numbers 25287039 and 15K17772, and also by 
the international program ``Climate And Weather of the Sun-Earth System - II (CAWSES-II): 
Towards Solar Maximum'' sponsored by SCOSTEP. This work was also supported by the 
``UCHUGAKU'' project of the Unit of Synergetic Studies for Space, Kyoto University.
AA is supported by Shiseido Female Researcher Science Grant.

{\it Facilities:} FMT, \facility{SDO (AIA)}, \facility{Hinode},
\facility{STEREO (SECCHI/EUVI), \facility{SOHO (MDI)}}


\clearpage

\end{document}